\documentclass[aps,prd,onecolumn,groupedaddress,showpacs,nofootinbib,amssymb
]{revtex4}
\usepackage[dvips]{graphicx}
\usepackage{amssymb}
\usepackage{amsmath}
\usepackage{graphicx,,color}
\usepackage{amsfonts}
\usepackage{bm}
\usepackage{cancel}
\usepackage{comment}

\allowdisplaybreaks[4]

\begin{document}

\tolerance=5000



\title{Can we learn from matter creation to solve the $H_{0}$ tension problem?}

\author{
Emilio Elizalde$^{1}$\footnote{E-mail: elizalde@ice.csic.es}, 
Martiros Khurshudyan$^{1}$\footnote{Email: khurshudyan@ice.csic.es},
Sergei D. Odintsov $^{1,2}$\footnote{Email: odintsov@ice.csic.es}
}
\affiliation{
$^{1}$ Institute of Space Sciences (ICE, CSIC) C. Can Magrans s/n, 08193 Barcelona, Spain\\
$^{2}$ ICREA, Passeig Luis Companys, 23, 08010 Barcelona, Spain}

\begin{abstract}

The $H_{0}$ tension problem is studied in the light of a matter creation mechanism (an effective approach to replacing dark energy), the way to define the matter creation rate being of pure phenomenological nature. Bayesian (probabilistic) Machine Learning is used to learn the constraints on the free parameters of the models, with the learning being based on the generated expansion rate, $H(z)$. Taking advantage of the method, the constraints for three redshift ranges are learned. Namely, for the two redshift ranges: $z\in [0,2]$~(cosmic chronometers) and $z\in [0,2.5]$~(cosmic chronometers + BAO), covering already available $H(z)$ data, to validate the learned results; and for a third redshift interval, $z\in[0,5]$, for forecasting purposes. It is learned that the $3\alpha H_{0}$ term in the creation rate provides options that have the potential to solve the $H_{0}$ tension problem.

\end{abstract}

\pacs {}

\maketitle

\section{Introduction}

After over twenty years since its discovery, the accelerated expansion of our universe is still in need of a consistent explanation \cite{Riess} - \cite{Hubble1}. A main task of modern cosmology is to validate some fundamental model where it could naturally fit. One of the first attempts to solve the problem was to interpret dark energy in terms of the cosmological constant. However, soon became clear that this specific dark energy model, although elegant and natural, has important issues~(such as the fine-tuning and cosmic coincidence problems), which generate, in some situations, tensions between the observed values and their theoretical estimates. To alleviate these problems with the cosmological constant, new dynamical dark energy models were soon developed, better supported by observational evidence. However, this provides, at best, just conditional or effective solutions, since additional problems appear. One of them, still unsolved to date, is the identification of the actual physics behind dark energy. It was later understood that a modification of General Relativity can also reasonably explain the accelerated expansion and it entangles various ideas about how to deal with the concept of dark energy (see \cite{Perivolaropoulos_LCDM} - \cite{GP_7}, and references therein). 

A modification of General Relativity performed at the action level allows dark energy to appear in the field equations in a most natural way, which avoids having to introduce it by hand. However, at the moment, it is still hard to say which of the above-mentioned approaches is the best, albeit this situation could change, in the future. More recently, another approach has been discussed, which also allows to effectively explain the accelerated expansion of the universe. In this approach, particle creation leads to the generation of a proper negative pressure to accelerate the universe's expansion (more details will be provided in Sect.~\ref{sec:FE}). In the recent literature,  interesting developments in such direction have appeared, some of which can be found in \cite{MC_start} - \cite{MC_end} (to mention just a few references).

The $H_{0}$ tension problem is one of the crucial issues in modern cosmology. The Planck CMB data analysis provides a value $H_{0} = 67.4 \pm 0.5$ km/s/Mpc, when the $\Lambda$CDM scenario is assumed~\cite{Plank}, while, a local measurement from the Hubble Space Telescope yields  $H_{0} = 73.52 \pm 1.62$ km/s/Mpc \cite{Hubble} (according to recent estimations, $H_{0} = 73.04 \pm 1.04$ km/s/Mpc \cite{Hubble1}) clearly indicating the existence of the tension problem. The tension between these $H_{0}$ estimations is notorious and surprising. It is key to understand if this discrepancy is due to new physics or if it just points to the fact that there is an overseen problem with the measurements or their comparison. The problem is hard since, as just mentioned, we still do not know the nature of dark energy,  not even that of dark matter. Possibly, a well-formulated dark energy model will contribute to the solution. In the recent literature, various possibilities in this respect have been put forward and some partial solutions have been obtained, but a fully convincing one is still lacking (see, for instance, \cite{Perivolaropoulos_LCDM} and \cite{Referee_1} and references therein).

The main idea in the present paper is to apply Machine Learning to find the $H_{0}$  value using a gravitationally induced matter creation mechanism. The goal is to learn first, from the procedure, the most reliable, unbiased form for the matter creation rate. This may allow to solve, or at least to alleviate, the $H_{0}$ tension problem. Our first results from the learning procedure look promising and indicate that the Bayesian Machine Learning approach is useful. Indeed, the method leads to a specific solution to the problem for cosmology with matter creation rate given by: $\Gamma_{c} = 3\alpha H_{0} + F(H)$, where $F(H)$ is a function of the universal expansion rate, $H$, only. The solution is learned using the generated expansion rate during the learning process, which makes the  Bayesian Machine Learning approach especially attractive. First,  the redshift ranges $z\in [0,2]$~(cosmic chronometers) and $z\in [0,2.5]$~(cosmic chronometers + BAO) will be considered, covering known $H(z)$ observations. Additionally, the models will be constrained using the generated expansion rate data for $z\in[0,5]$, just for forecasting purposes. At the end of the present study, available $H(z)$ observational data will be used to validate the learned results (see Table \ref{tab:Table0}). 

We should stress that, recently, Bayesian (Probabilistic) Machine Learning has been applied to tackle the $H_{0}$ tension problem, but just with the help of a single inhomogeneous viscous universe model \cite{Elizalde_9}. Another study has shown that the method can be used to constrain the cosmic opacity value, by revealing its connection with the $H_{0}$ tension problem~\cite{Elizalde_8}. And,  in \cite{Elizalde_7}, the authors found that also the Swampland criteria can constrain valid models for a dark energy-dominated universe. All such results have been obtained using Bayesian (Probabilistic) Machine Learning, and already provide hints that the $H_{0}$ tension problem may be due to a deviation from the cold dark matter paradigm. In other words, it has been pointed out with these methods that, on cosmological scales, dark matter might not be that cold, and that this could be the reason why the proposed interacting dark energy models work in cosmology \cite{Elizalde_H0}. On the other hand, another recent paper using a Gaussian Process has confirmed that deviation from the cold dark matter model can provide a solution to the $H_{0}$ tension problem \cite{Elizalde_H0_recent}.\\ 

This work is organized as follows. In Sect.~\ref{sec:FE}, the assumptions leading to the background dynamics are listed, which will be used for expansion rate generation and in the learning process. We stress, once more, that the learning procedure is applied to the generated data and that later the learned constraints are used for comparison with the existing observational data~(see Table~\ref{tab:Table0}). The essential aspects behind the Bayesian (Probabilistic) Machine Learning approach are discussed in Sect.~\ref{sec:method}. The results obtained are commented in Sect.~\ref{sec:Res}, which is followed by an analysis of their implications, relevant for the $H_{0}$ tension and other issues. The conclusions of the present analysis are to be found in Sect.~\ref{sec:conc}. 

\begin{table}[t]
  \centering
    \begin{tabular}{ |  l   l   l  |  l   l  l  | p{2cm} |}
    \hline
$z$ & $H(z)$ & $\sigma_{H}$ & $z$ & $H(z)$ & $\sigma_{H}$ \\
      \hline
$0.070$ & $69$ & $19.6$ & $0.4783$ & $80.9$ & $9$ \\
         
$0.090$ & $69$ & $12$ & $0.480$ & $97$ & $62$ \\
    
$0.120$ & $68.6$ & $26.2$ &  $0.593$ & $104$ & $13$  \\
 
$0.170$ & $83$ & $8$ & $0.680$ & $92$ & $8$  \\
      
$0.179$ & $75$ & $4$ &  $0.781$ & $105$ & $12$ \\
       
$0.199$ & $75$ & $5$ &  $0.875$ & $125$ & $17$ \\
     
$0.200$ & $72.9$ & $29.6$ &  $0.880$ & $90$ & $40$ \\
     
$0.270$ & $77$ & $14$ &  $0.900$ & $117$ & $23$ \\
       
$0.280$ & $88.8$ & $36.6$ &  $1.037$ & $154$ & $20$ \\
      
$0.352$ & $83$ & $14$ & $1.300$ & $168$ & $17$ \\
       
$0.3802$ & $83$ & $13.5$ &  $1.363$ & $160$ & $33.6$ \\
      
$0.400$ & $95$ & $17$ & $1.4307$ & $177$ & $18$ \\

$0.4004$ & $77$ & $10.2$ & $1.530$ & $140$ & $14$ \\
     
$0.4247$ & $87.1$ & $11.1$ & $1.750$ & $202$ & $40$ \\
     
$0.44497$ & $92.8$ & $12.9$ & $1.965$ & $186.5$ & $50.4$ \\

$$ & $$ & $$ & $$ & $$ & $$\\ 

$0.24$ & $79.69$ & $2.65$ & $0.60$ & $87.9$ & $6.1$ \\
$0.35$ & $84.4$ & $7$ &  $0.73$ & $97.3$ & $7.0$ \\
$0.43$ & $86.45$ & $3.68$ &  $2.30$ & $224$ & $8$ \\
$0.44$ & $82.6$ & $7.8$ &  $2.34$ & $222$ & $7$ \\
$0.57$ & $92.4$ & $4.5$ &  $2.36$ & $226$ & $8$ \\ 
          \hline
    \end{tabular}
    \vspace{5mm}
\caption{$H(z)$ and its uncertainty $\sigma_{H}$  in  units of km s$^{-1}$ Mpc$^{-1}$. The upper panel consists of thirty samples deduced from the differential age method. The lower panel corresponds to ten samples obtained from the radial BAO method. The table is constructed according to \cite{HTable_0} - \cite{HTable_13}. The data was only used to validate the results obtained from Bayesian Machine Learning. It was not directly used in the learning process and thus did not have a direct impact on the results we achieved.}
  \label{tab:Table0}
\end{table}

\section{Cosmology with matter creation}\label{sec:FE}

In this section, we briefly discuss the field equations describing the cosmological background evolution with matter creation. In particular,  a gravitationally induced particle creation scenario is considered, which can replace dark energy. However, even in this case, when we do not need to introduce dark energy explicitly, we still have to consider a parameterization for the particle creation rate. This means that, even within this scenario, which provides an interesting modification of General Relativity, we cannot fully avoid phenomenology. In the recent literature, there is an intensive study of cosmological models with different matter creation rates successfully addressing the issue of the accelerated expansion of the universe\footnote{see, for instance, more recent work \cite{MC_10} and references therein along with \cite{MC_Ref_1} and \cite{MC_Ref_2}  to better follow the discussion in this section.}. We here consider a homogeneous and isotropic flat Friedmann-Lemaître-Robertson-Walker (FLRW) model, 
\begin{equation}\label{eq:FRW}
ds^{2} = -dt^{2} + a^{2}(t) \left [ dr^{2} + r^{2} (d \theta^{2} + \sin^{2} \theta d \phi^{2} ) \right ],
\end{equation}
where $a(t)$ is the scale factor, to be used to obtain the final form of the field equations employed in our study. We start from the Einstein field equations, given by
\begin{equation}\label{eq:EE}
R_{\mu\nu} - \frac{1}{2}R g_{\mu\nu} = T_{\mu \nu},
\end{equation}
where $T_{\mu\nu}$ is the energy-momentum tensor describing the matter content of the Universe
\begin{equation}
T_{\mu\nu} = (\rho + P) u_{\mu} u_{\nu} + P g_{\mu \nu},
\end{equation}
empowered by the matter creation mechanism and satisfying the covariant conservation equation $T^{\mu\nu} ; \nu = 0$. In the last two equations, $R_{\mu\nu}$, $R$, $g_{\mu\nu}$ and $u_{\mu}$ are the Ricci curvature tensor, the scalar curvature, the metric tensor, and the fluid four-velocity, respectively. Correspondingly, $\rho$ and $P$ are the energy density and the pressure of the energy source. Moreover, $P = P + P_{c}$, where $P$ is the equilibrium pressure and $P_{c}$ is the pressure due to matter creation. It is well known that in the homogeneous and isotropic flat FLRW model, the dissipative phenomenon may be in the form of bulk viscous pressure either due to the coupling of different components of the cosmic substratum or due to the non-conservation of the particle number. Therefore, for an open thermodynamical system where the number of fluid particles is not preserved, the particle conservation equation gets modified and adopts the following form
\begin{equation}\label{eq:CONVEQ}
N^{\mu}_{;\mu} = n_{,\mu}u^{\mu} + 3Hn = n\Gamma_{c},
\end{equation}       
where $\Gamma_{c}$ is the particle creation rate, while $N^{\mu} = nu^{\mu}$ is the particle flow vector. Moreover, $n=N/V$ is the particle number density in a comoving volume $V$ containing $N$ particles. Now, if we use the conservation equation, Eq.~(\ref{eq:CONVEQ}), from the Gibb's relation
\begin{equation}
Tds = d \left( \frac{\rho}{n}\right ) + P d\left( \frac{1}{n}\right),
\end{equation}
we can get the variation of the entropy per particle
\begin{equation}\label{eq:VENT}
nT\dot{s} = \dot{\rho} + 3H \left( 1 - \frac{\Gamma_{c}}{3H}\right) (\rho + P).
\end{equation} 
In Eq.~(\ref{eq:VENT}), $T$ is the fluid temperature, and $s$ is the entropy per particle. Moreover, if we assume the thermodynamic system to be an ideal one, i.e. $\dot{s}=0$, then from Eq.~(\ref{eq:VENT}) we obtain the conservation equation, as
\begin{equation}
\dot{\rho} + 3H(\rho + P) = \Gamma_{c}(\rho + P).
\end{equation}
The last equation can be re-written, namely 
\begin{equation}
\dot{\rho} + 3H(\rho + P + P_{c}) = 0,
\end{equation}
where $P_{c}$ is termed as the particle creation pressure and reads 
\begin{equation}
P_{c} = -\frac{\Gamma_{c}}{3H}(\rho + P).
\end{equation}
In our learning process, we neglect radiation and baryons, assuming that the main constituent of the universe is cold dark matter and that there is only cold dark matter particle creation. Therefore, we arrive to the following set of equations~(with $8\pi G = c = 1$)
\begin{equation}\label{eq:FEQ1}
H^{2} = \frac{1}{3} \rho,
\end{equation}
\begin{equation}\label{eq:FEQ2}
\dot{H} = -\frac{1}{2}( \rho + P_{c} ),
\end{equation}
with 
\begin{equation}\label{eq:Pc}
P_{c} = -\frac{\Gamma_{c}}{3H}\rho.
\end{equation}
to describe the background dynamics, where the dot means differentiation with respect to the cosmic time. The nature of the particle creation rate, $\Gamma_{c}$, is not known yet and, in general, there is no specific bound that could allow us to choose some particular forms for it. The only way to decide whether the form chosen for the creation rate is viable or not is to involve observational constraints on the cosmological model. The details of the approach used in this work will be discussed in the next section.

\section{Methodology}\label{sec:method}

In this section, the building blocks of Bayesian Machine Learning will be discussed, allowing us to explore the parameter space of the models in the presence of a gravitationally induced particle creation mechanism.

\subsection{Bayesian Machine Learning}

To estimate whether the model is applicable or not, we need to use data and constrain the given cosmological model. Markov chain Monte Carlo (MCMC) and similar Bayesian modeling approaches play a central role in this chain allowing also to reduce the phenomenology. Here, first, we provide a basic motivation to clarify why it is desirable to look for alternative ways to constrain models. In other words, we want to justify why we consider replacing MCMC.  The Bayes theorem
\begin{equation}\label{eq:BT}
    P(\theta | \mathcal{D}) = \frac{P(\mathcal{D} |\theta) P(\theta)}{P(\mathcal{D})},
\end{equation}
allows to find the probability of a given model with $\theta$ parameters explaining given data $\mathcal{D}$ (a conditional probability). In the above equation, $P(\theta)$ is the prior belief on the parameter $\theta$ describing the model at hand. $P(\mathcal{D}|\theta)$ is the likelihood representing the probability of observing the data $\mathcal{D}$ given parameter $\theta$, while $P(\mathcal{D})$ is the marginal likelihood or model evidence and it is useful for the model selection process. The computation of the probabilities is always a hard problem. Therefore, developing alternative methods to overcome the associated difficulties is a must. The variational inference discussed below is one of the alternative methods for performing Bayesian inference. It is very attractive for cosmological and astrophysical applications, because it is considerably faster than MCMC techniques. Variational inference suggests solving an optimization problem by approximating the target probability density. Here, to measure such proximity the Kullback-Leibler (KL) divergence is used \cite{KL}. The case we are interested in the target probability density is the Bayesian posterior, from where one can infer the constraints on the model parameters. In this algorithm, the first step is finding a family of densities $\mathcal{Q}$ and corresponding member of that family $q(\theta) \in \mathcal{Q}$, which is the closest one to the target probability density. The member thus found minimizes the KL  divergence to the exact posterior
\begin{equation}
q^{*}(\theta)=\underset{q(\theta) \in \mathcal{Q}}{\arg \min } \mathrm{KL}(q(\theta)|p(\theta|\mathcal{D})),
\end{equation}
 where $\theta$ is the latent variable to measure of such proximity, and the KL divergence is defined as
\begin{equation}\label{eq:KL1}
\mathrm{KL}(q(\theta) \| p(\theta \mid  \mathcal{D}))=\mathbb{E}_{q(\theta)}[\log q(\theta)]-\mathbb{E}_{q(\theta)}[\log p(\theta \mid  \mathcal{D})].
\end{equation}
Bayes theorem allows us to rewrite the above KL divergence as   
\begin{equation}\label{eq:KL2}
\mathrm{KL}(q(\theta) \| p(\theta \mid  \mathcal{D}))=\log p( \mathcal{D})+\mathbb{E}_{q(\theta)}[\log q(\theta)]-\mathbb{E}_{q(\theta)}[\log p( \mathcal{D}, \theta)].
\end{equation}
Therefore, to minimize the above KL divergence term, one needs to minimize the second and third terms in Eq.(\ref{eq:KL2}). On the other hand, the variational  lower bound can be rewritten as \cite{ELBO} 
\begin{equation}\label{eq:ELBO}
\operatorname{ELBO}(q(\theta))=\mathbb{E}_{q(\theta)}[\log p( \mathcal{D} \mid \theta)]-\mathrm{KL}(q(\theta) \| p(\theta)).
\end{equation}
The first term is a sort of data fit term maximizing the likelihood of the observational data. The KL-divergence between the variational distribution and the prior is given by the second term 
in the above equation. It is a regularization term controlling the variational distribution, because we need it not to become too complex and to allow us to avoid over-fitting. With this, we wish to stress that we get a fitting tool to control the process and efficiently avoid computation problems by increasing or decreasing the contribution of the first two terms in Eq.(\ref{eq:KL1}). Various interesting and valuable information on this topic can be found, e.g., in \cite{51} - \cite{54}. 

Still to be discussed is how to find the approximation for the target probability density. In what follows, we will learn it, from the considered cosmological model, directly using Neural Networks (NN). Now, the terms in Eq.(\ref{eq:ELBO}) can be reduced to the initial personal belief and the generated model belief, respectively. This is a convenient approach, because now even with low-quality data one can validate the learned results. Deep probabilistic learning has been used as a learning method. It is a type of deep learning accounting for uncertainties in the model, initial belief, belief update, and deep neural networks. This approach provides adequate groundwork to output reliable estimations for many ML tasks.

\subsection{Implementation of Bayesian Machine Learning}

We will here use the probabilistic programming package \textit{PyMC3} \cite{PyMC3}, based on the deep learning library Theano. This is a deep learning Python-based library, very useful to perform variational inference and to build the posterior distribution. On passing, we strongly suggest that readers interested in exploring Bayesian Machine Learning and variational inference follow the examples and discussions provided in the \textit{PyMC3} manual.

Now, let us discuss how we should understand the above discussion, which allows us to integrate Bayesian Machine Learning and variational inference to learn the constraints on the crafted cosmological models we consider. To be short, in our analysis using PyMC3, we have followed the steps:

\begin{enumerate}

\item Define the model to be used to generate missing or low-quality observational data. In our case, it will be the cosmological models with $\Gamma_{c}$ matter creation rate. The background dynamics of this model have been discussed in Sect.~\ref{sec:FE} and the associated observable has been taken to be the expansion rate of the universe. 

\item Envisage the data used in the learning process to be the data obtained from the generative process. As we mentioned above, the observational data will be the generated expansion rate of the universe.  Then, we generate probability distributions showing how a given cosmological model can explain the data. In this way, the complexity of the problem will be significantly reduced, because the family of probabilities approximating the final posterior will directly depend only on priors. To notice dependency, we need to consider the meaning of the right-hand side of the Bayes theorem, Eq.(\ref{eq:BT}). 

\item Run the learning algorithm, to get a brand new distribution over the model parameters and update prior beliefs. This indicates that we need the prior distributions over the model's free parameters indicating the initial belief. Then they will be updated to get the posteriors. One very crucial aspect appearing here is the possibility of learning the true prior belief imposed on the model's free parameters. Generating the learning process and the direction of the learning is always controlled by the KL divergence. Since we have involved probabilistic programming, after enough generated probabilistic distributions, we expect to learn the asymptotically correct form for the posterior distribution allowing us to infer the constraints.  

\end{enumerate}

To end this section, we should stress that starting from different initial beliefs, our learning procedure asymptotically converges to the results discussed in this paper.

\section{Study of models with different matter creation rates}\label{sec:Res}

In this section, we present and discuss our learned results. In our learning procedure, the generated data is the expansion rate of the universe built from Eqs.~(\ref{eq:FEQ1}), (\ref{eq:FEQ2}), and~(\ref{eq:Pc}), supplied by the form of the matter creation rates discussed in this section. We have considered three $z\in [0,2.0]$, $z\in [0,2.5]$, and $z \in [0,5]$ redshift ranges to understand how our learned knowledge can change if we use generated expansion rate data. At this very early stage of the analysis, we did not consider other generated observables. We wanted to learn a solution for the $H_{0}$ tension problem and avoid any artificial bias that could emerge when other generated observational datasets are used either together, or separately. Of course, this could be done, as soon as the bias of any sort is translated into proper probability distributions, which will also enter into the learning process. In general, if the basic characteristics of the model are well understood, then it is very straightforward to extend the analysis to include other datasets, too, taking into account the bias. Among other issues, this one has been left to be more appropriately considered in forthcoming papers.

\subsection{Model with $\Gamma_{c} = 3 \beta H_{0} (H(z)/H_{0})^{\alpha}$}

\begin{figure}[h]
 \begin{center}$
 \begin{array}{cccc}
\includegraphics[width=120 mm]{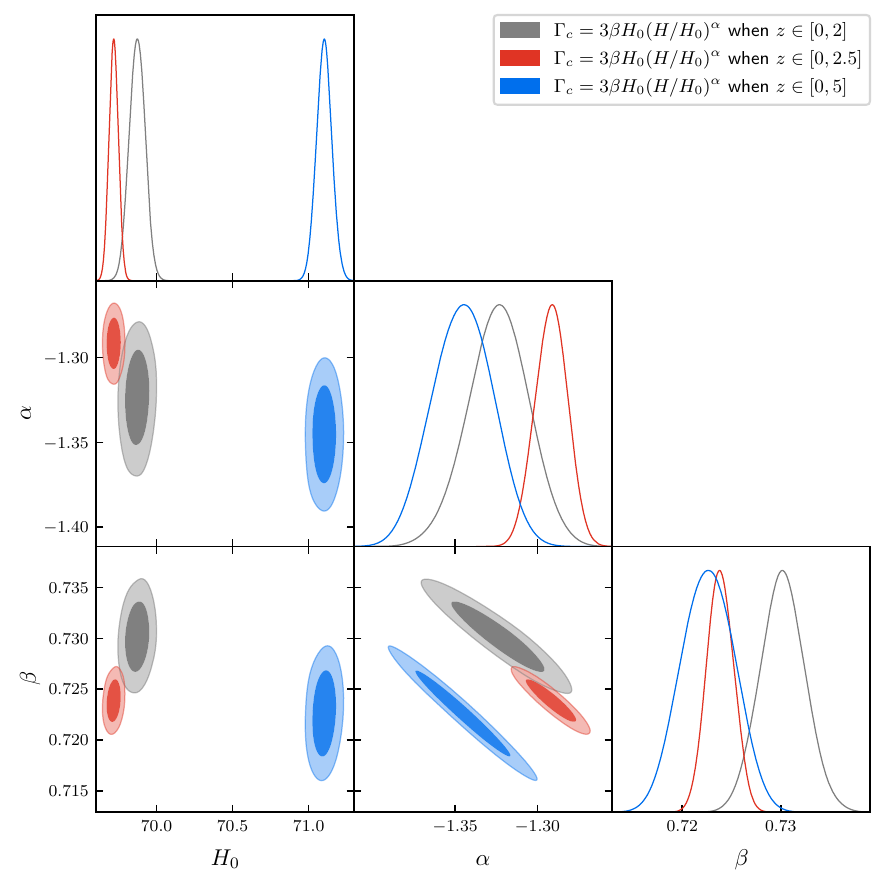}
 \end{array}$
 \end{center}
\caption{Contour maps of the model given by Eq.~(\ref{eq:Gc1}), for $z \in [0,2.0]$, $z \in [0,2.5]$, and $z \in [0,5]$, respectively. The best-fit values of the model parameters are presented in Table~\ref{tab:Table1}. In all three cases, $H_{0}\in[65.5,85.5]$, $\alpha \in [-1.5, 1.5]$, and $\beta \in [-0.5,0.5]$, flat priors have been imposed to generate the ``observational'' expansion rate data from Eq.~(\ref{eq:HGcOdintsov}). The analysis is based on 10 chains and in each chain, 15,000 ``observational'' data sets have been simulated/generated.}
 \label{fig:Fig0_Odintsov}
\end{figure}

The first model we analyzed is given by the following matter creation rate (see for instance \cite{MC_10})
\begin{equation}\label{eq:GcOdintsov}
\Gamma_{c} = 3 \beta H_{0} \left ( \frac{H(z)}{H_{0}} \right )^{\alpha},
\end{equation}
where $H_{0}$, $\alpha$ and $\beta$ are the model's free parameters to be learned, while $H(z)$ is the expansion rate. It is not hard to see that this is a model, where 
\begin{equation}\label{eq:HGcOdintsov}
H^{\prime}(z) = \frac{3 H(z)}{2} \left ( \frac{1-\beta  H^{1-\alpha}_{0} H(z)^{-1 + \alpha }}{1+z} \right ).
\end{equation}
It should be mentioned that have kept the same notations as in \cite{MC_10}, to make the comparison of results easier. We see that learned results, for the mean values of the parameters, are in good agreement with previous results. However, here, the learned errors are smaller, indicating the possibility of getting very tight constraints \cite{MC_10}. The learned constraints can be found in Table \ref{tab:Table1_Odintsov}, and the learned contour maps of this model can be found in Fig.~(\ref{fig:Fig0_Odintsov}). We see how the constraints on the model parameters are changing when the considered redshift ranges change. Moreover, using Bayesian (Probabilistic) Machine Learning approach, we learned that the model is not able to solve the $H_{0}$ tension problem. 
 
\begin{table}
  \centering
    \begin{tabular}{ | c | c | c | c | c | c |  p{2cm} |}
    \hline
    
 $\Gamma_{c} = 3 \beta H_{0}  (H(z)/H_{0})^{\alpha}$ & $H_{0}$ & $\alpha$ & $\beta$ \\
      \hline
 
  when $z\in[0,2]$ & $69.87 \pm 0.15$ km/s/Mpc & $-1.323 \pm 0.018$ & $0.7302 \pm 0.0023$\\
          \hline
          
 when $z\in[0,2.5]$ & $69.72 \pm 0.03$ km/s/Mpc & $-1.292 \pm 0.011$ & $0.7239 \pm 0.0014$ \\
          \hline
          
when $z\in[0,5]$  & $71.11 \pm 0.05$ km/s/Mpc& $-1.345 \pm 0.019$ & $0.7226 \pm 0.0028$ \\

           \hline
 
     \end{tabular}
\caption{Best fit values and $1\sigma$ errors estimated for the model given by Eq.~(\ref{eq:GcOdintsov}), when $z \in [0,2.0]$, $z \in [0,2.5]$, and $z \in [0,5]$, respectively. The results have been obtained from the Bayesian (Probabilistic) Machine Learning approach, where the generative-based process relies on Eq.~(\ref{eq:HGcOdintsov}), using $H_{0}\in[65.5,85.5]$, $\alpha \in [-2.5, 2.5]$ and $\beta \in [-1.5,1.5]$ as flat priors. The analysis is based on 10 chains and in each chain, 15,000 ``observational'' data sets have been simulated/generated.}
  \label{tab:Table1_Odintsov}
\end{table}

It should be mentioned that, in \cite{MC_10}, new interesting solutions assuming that $\Gamma_{c}$ can be a function of $H$ and its higher order derivatives have been obtained, too. The analysis of such models is not our goal here and is left to forthcoming papers. In the present one, we will consider only models where $\Gamma_{c} = \Gamma(H)$. To end the discussion here, we wish to indicate that a reanalysis of the specific model from \cite{MC_10}  allowed us once again to validate the robustness of the adopted approach.

\subsection{Model with $\Gamma_{c} = 3 \alpha H_{0} + 3 \beta H(z)$}

The first model we have considered, where matter creation rate $\Gamma_{c}$ is only a function of the expansion rate,  is based on the following assumption
\begin{equation}\label{eq:Gc1}
\Gamma_{c} = 3 \alpha H_{0} + 3 \beta H(z),
\end{equation}
where $H_{0}$, $\alpha$ and $\beta$ are the model free parameters to be learned, while $H(z)$ is the expansion rate. Starting from Eqs.~(\ref{eq:FEQ1}) and (\ref{eq:FEQ2}),  and taking into account Eqs.~(\ref{eq:Pc}) and (\ref{eq:Gc1}),  after some algebra, we can write the expansion rate of the universe in the following way\footnote{Actually, we have obtained two solutions, namely $H = 0$ and the one given by Eq.~(\ref{eq:HGc1}).}

\begin{equation}\label{eq:HGc1}
H(z) = \frac{\alpha H_{0}- H_{0} (\alpha +\beta -1) (z+1)^{-\frac{3}{2} (\beta -1)}}{1-\beta }.
\end{equation}

\begin{table}
  \centering
    \begin{tabular}{ | c | c | c | c | c | c |  p{2cm} |}
    \hline
    
 $\Gamma_{c} = 3 \alpha H_{0} + 3 \beta H(z)$ & $H_{0}$ & $\alpha$ & $\beta$ \\
      \hline
 
  when $z\in[0,2]$ & $66.53 \pm 0.15$ km/s/Mpc & $0.4201 \pm 0.0077$ & $0.025 \pm 0.005$\\
          \hline
          
 when $z\in[0,2.5]$ & $67.43 \pm 0.16$ km/s/Mpc & $0.4362 \pm 0.0079$ & $0.025 \pm 0.005$ \\
          \hline
          
when $z\in[0,5]$  & $66.77 \pm 0.16$ km/s/Mpc& $0.4160 \pm 0.0024$ & $0.0027 \pm 0.001$ \\

           \hline
 
     \end{tabular}
\caption{Best fit values and $1\sigma$ errors estimated for the model given by Eq.~(\ref{eq:Gc1}), when $z \in [0,2.0]$, $z \in [0,2.5]$ and $z \in [0,5]$, respectively. The results have been obtained from the Bayesian (Probabilistic) Machine Learning approach, where the generative-based process relies on Eq.~(\ref{eq:HGc1}), using $H_{0}\in[65.5,85.5]$, $\alpha \in [-1.5, 1.5]$ and $\beta \in [-0.5,0.5]$ as flat priors. The analysis is based on 10 chains and in each chain, 15,000 ``observational'' data sets have been simulated/generated.}
  \label{tab:Table1}
\end{table}

\begin{figure}[h]
 \begin{center}$
 \begin{array}{cccc}
\includegraphics[width=120 mm]{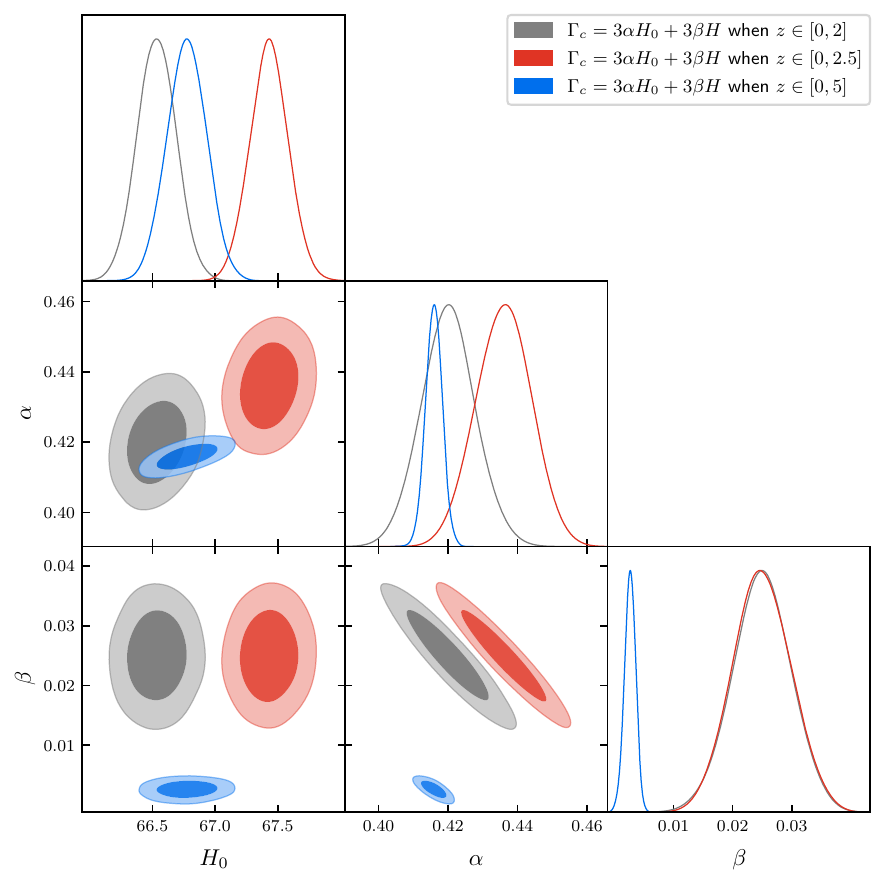}
 \end{array}$
 \end{center}
\caption{Contour maps of the model given by Eq.~(\ref{eq:Gc1}) for $z \in [0,2.0]$, $z \in [0,2.5]$ and $z \in [0,5]$, respectively. The best-fit values of the model parameters are presented in Table~\ref{tab:Table1}. In all three cases, $H_{0}\in[65.5,85.5]$, $\alpha \in [-1.5, 1.5]$ and $\beta \in [-0.5,0.5]$ flat priors have been imposed to generate the ``observational'' expansion rate data from Eq.~(\ref{eq:HGc1}). The analysis is based on 10 chains and in each chain, 15,000 ``observational'' data sets have been simulated/generated.}
 \label{fig:Fig0_1_a}
\end{figure}

This is the form of the expansion rate used in our generative process allowing us to do the learning. Imposing $H_{0}\in[65.5,85.5]$, $\alpha \in [-1.5, 1.5]$ and $\beta \in [-0.5,0.5]$ flat priors on the free parameters, we used $10$ chains to generate $15,000$ "observational" data-sets in each to do the learning. The learned constraints can be found in Table~\ref{tab:Table1}. In particular, for this model, we have learned that:
\begin{itemize}

\item The best-fit values for the model free parameters with $1\sigma$ error are $H_{0} = 66.53 \pm 0.15$ km/s/Mpc, $\alpha =0.4201 \pm 0.0077$ and $\beta = 0.025 \pm 0.005$, when $z \in [0,2.0]$. The contour map is given in Fig.~(\ref{fig:Fig0_1_a}), in grey color. 

\item On the other hand, when $z \in [0,2.5]$, the most likely best fit values with $1\sigma$ error are: $H_{0} = 67.43 \pm 0.16$ km/s/Mpc, $ \alpha =0.4362 \pm 0.0079$ and $\beta = 0.025 \pm 0.005$. The contour map is given in Fig.~(\ref{fig:Fig0_1_a}), in red color. 

\item Finally, when $z \in [0,5.0]$, the most likely best fit values with $1\sigma$ error are: $H_{0} = 66.77 \pm 0.16$ km/s/Mpc, $ \alpha = 0.4160 \pm 0.0024$ and $\beta = 0.0027 \pm 0.001$. The contour map is given in Fig.~(\ref{fig:Fig0_1_a}), in blue color. 

\end{itemize}

\begin{figure}[t!]
 \begin{center}$
 \begin{array}{cccc}
\includegraphics[width=80 mm]{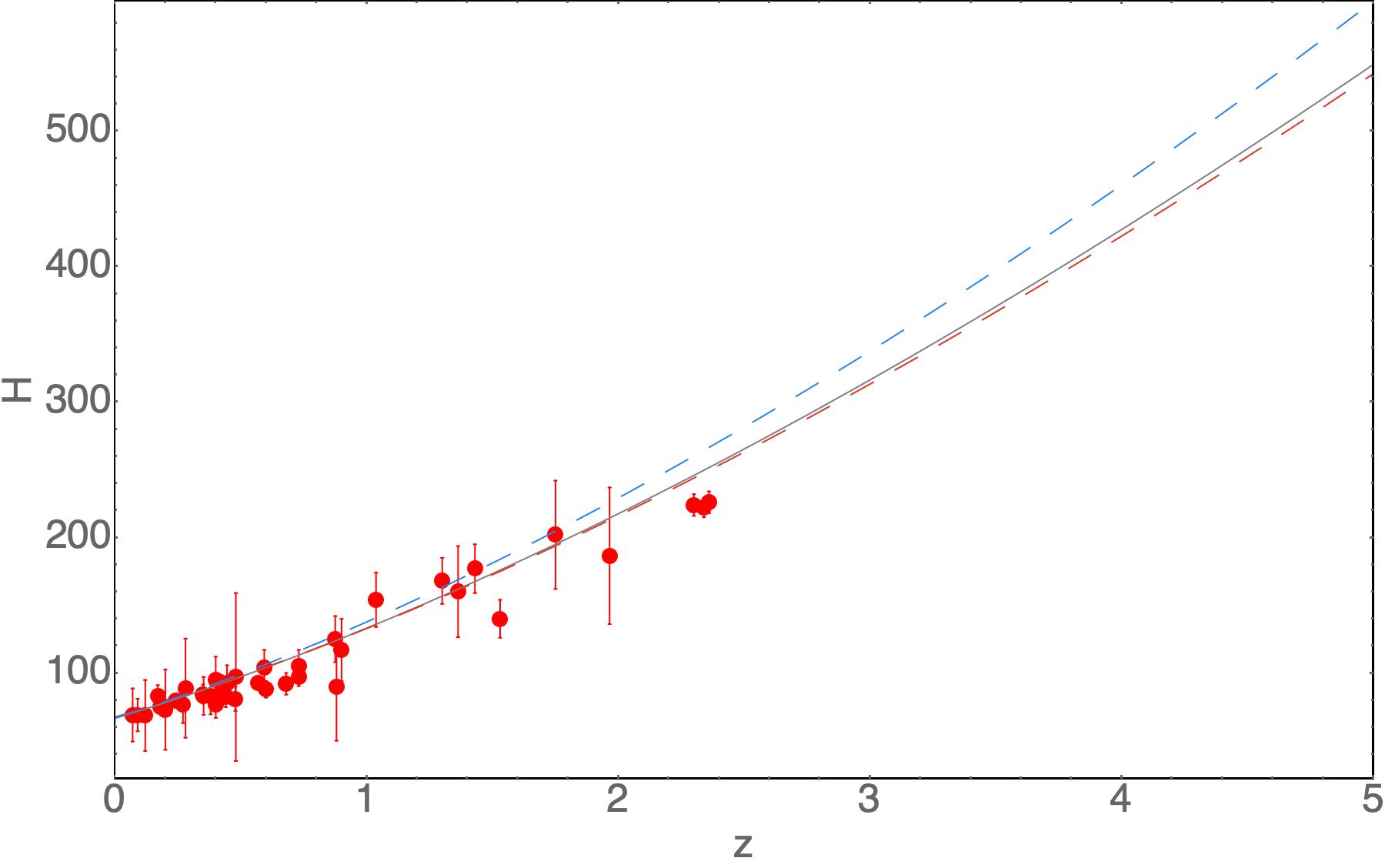}&&
\includegraphics[width=80 mm]{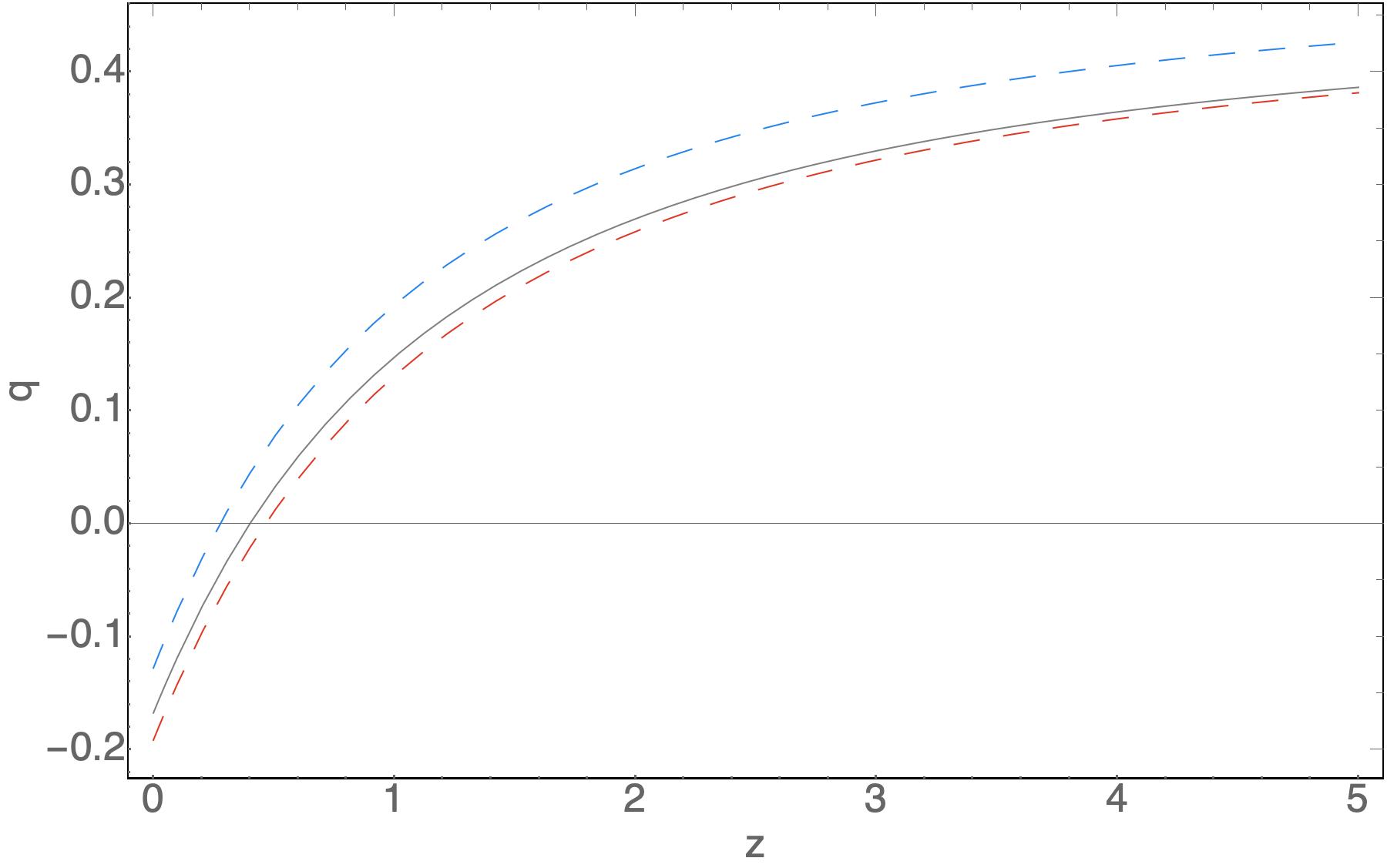}\\
 \end{array}$
 \end{center}
\caption{The behavior of the Hubble parameter, in comparison with known $H(z)$ data, is presented on the left-hand side of the plot. The grey curve is a plot of the Hubble parameter for the best-fit values of the model parameters when $z \in [0,2]$; the dashed red curve corresponds to the case when $z \in [0,2.5]$; while the dashed blue curve stands for the case when $z \in [0,5]$. The red dots represent the known observational $H(z)$ data as in Table~\ref{tab:Table0}. The right-hand side depicts the behavior of the deceleration parameter $q(z)$. The model is given by Eqs.~(\ref{eq:Gc1}) and (\ref{eq:HGc1}). In both cases, only the best-fit values for the model-free parameters obtained by the Bayesian Learning approach and presented in Table~\ref{tab:Table1} have been used.}
 \label{fig:Fig0_2_a}
\end{figure}

Therefore, from the learned constraints, we can conclude that this model cannot solve the $H_{0}$ tension issue. Moreover, we found that the $H_{0}$ is very sensitive to the learning redshift range. In particular, we learned that the mean value of $H_{0}$, when $z\in [0,2.5]$, is bigger than what we learned when $z\in [0,2]$. On the other hand, when $z\in [0,5]$, the forecast $H_{0}$ mean is smaller than the one found when $z\in [0,2.5]$. A similar result has been learned for the $\alpha$ parameter. There is an interesting situation with the parameter $\beta$, allowing to conclude that it will remain unchanged for both redshift ranges: $z\in [0,2]$ and $z\in [0,2.5]$. However, when $z\in [0,5]$ its mean value will decrease significantly: from $\beta = 0.025$ to $\beta = 0.0027$, respectively.

Besides the fact that the model cannot solve the $H_{0}$ tension problem, we found a hint that the model cannot either explain the high-redshift expansion rate data. There is a huge tension between the high-redshift expansion rate data and the learned theoretical results. Therefore, this model should be rejected. We already mentioned that the entire learning process is based on the generated expansion rate data and that we use available observational data to validate our learned results. As we can see, Fig.~(\ref{fig:Fig0_2_a}) clearly indicates that the model is not valid and, most likely, future high redshift expansion rate data will not allow the model to be reconsidered. The right-hand side plot in Fig.~(\ref{fig:Fig0_2_a}) depicts the behavior of the deceleration parameter $q(z)$, indicating that the model can explain the accelerated expansion of the universe. 

To finish, we should mention that, despite the model can explain the accelerated expansion, it cannot actually solve the $H_{0}$ tension problem, and it is also in tension with the high-redshift expansion rate data. Therefore, it should be rejected, too. However, it can be used to reveal a bias, in the case, other observational datasets, used within MCMC analysis, indicate a valid solution to one of the mentioned problems. We remind the reader that, in the last situation, we would need to evaluate the likelihood and then one is limited to the ranges where observational data are available, only.

\subsection{Model with $\Gamma_{c} = 3 \alpha H_{0} + 3 \gamma H(z) \cos(1+z)$ }

\begin{table}
  \centering
    \begin{tabular}{ | c | c | c | c | c | c |  p{2cm} |}
    \hline
    
$\Gamma_{c} = 3 \alpha H_{0} + 3 \gamma  \cos(1+z) H(z)$ & $H_{0}$ & $\alpha$ & $\gamma$ \\
      \hline
 
  when $z\in[0,2]$ & $68.85 \pm 0.15$ km/s/Mpc & $0.571 \pm 0.002$ & $0.197 \pm 0.012$\\
          \hline
          
 when $z\in[0,2.5]$ & $67.82 \pm 0.15$ km/s/Mpc & $0.569\pm 0.002$ & $0.215 \pm 0.007$ \\
          \hline
          
when $z\in[0,5]$  & $68.03 \pm 0.15$ km/s/Mpc& $0.574 \pm 0.002$ & $0.229 \pm 0.003$ \\

           \hline    
 
     \end{tabular}
\caption{Best fit values and $1\sigma$ errors estimated for the model given by Eq.~(\ref{eq:Gc2}), for $z \in [0,2.0]$, $z \in [0,2.5]$ and $z \in [0,5]$, respectively. The results have been obtained from our Bayesian (Probabilistic) Machine Learning approach, where the generative-based process is based on Eq.~(\ref{eq:HGc2}) using $H_{0}\in[65.5,85.5]$, $\alpha \in [-1.5, 1.5]$ and $\gamma \in [-0.5,1.5]$ flat priors. The analysis is based on 10 chains and in each chain, 15,000 ``observational'' data sets have been simulated/generated.}
  \label{tab:Table2}
\end{table}

The second model relies on the assumption that the function $F(H)$ in $\Gamma_{c} = 3\alpha H_{0} + F(H)$ is given by $3 \gamma \cos(1+z)  H(z)$. In this case, the matter creation rate adopts the following form
\begin{equation}\label{eq:Gc2}
\Gamma_{c} = 3 \alpha H_{0} + 3 \gamma \cos(1+z) H(z) ,
\end{equation}
with $H_{0}$, $\alpha$ and $\gamma$ free parameters to be learned. Starting from Eqs.~(\ref{eq:FEQ1}) and (\ref{eq:FEQ2}),  and taking into account Eqs.~(\ref{eq:Pc}) and (\ref{eq:Gc2}), after some algebra, we derive the expansion rate dynamics of the universe, as follows
\begin{equation}\label{eq:HGc2}
H^{\prime}(z) = -\frac{3 (\gamma  H(z) \cos (z+1)-H(z)+\alpha H_{0})}{2 (z+1)},
\end{equation}
where the prime stands for the derivative with respect to the redshift. Again, imposing $H_{0}\in[65.5,85.5]$, $\alpha \in [-1.5, 1.5]$ and $\gamma \in [-0.5,1.5]$ as flat priors on the free parameters, we have used $10$ chains to generate $15,000$ ``observational'' data-sets in each to do the learning of the constraints. For convenience, the learned constraints have been collected in Table~\ref{tab:Table2}. From the results, we conclude that

\begin{figure}[h!]
 \begin{center}$
 \begin{array}{cccc}
\includegraphics[width=120 mm]{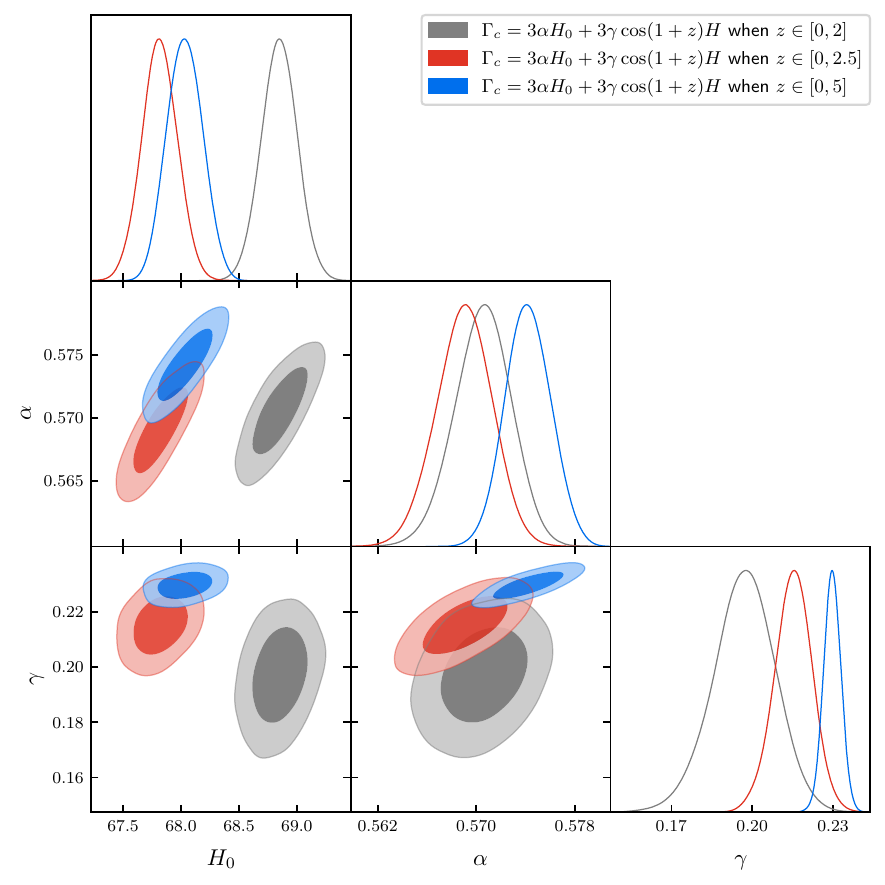}
 \end{array}$
 \end{center}
\caption{Contour maps of the model given by Eq.~(\ref{eq:Gc2}) for $z \in [0,2.0]$, $z \in [0,2.5]$ and $z \in [0,5]$, respectively. The best-fit values of the model parameters are shown in Table~\ref{tab:Table2}. In all three cases, $H_{0}\in[65.5,85.5]$, $\alpha \in [-1.5, 1.5]$ and $\gamma \in [-0.5,1.5]$ flat priors have been imposed to generate ``observational'' expansion rate data from Eq.~(\ref{eq:HGc2}). The analysis is based on 10 chains and in each chain, 15,000 ``observational'' data sets have been simulated/generated.}
 \label{fig:Fig0_1_b}
\end{figure}

\begin{itemize}

\item The best-fit values for the model free parameters with $1\sigma$ error are $H_{0} = 68.85 \pm 0.15$ km/s/Mpc, $\alpha = 0.571 \pm 0.002$ and $\gamma = 0.197 \pm 0.012$, when $z \in [0,2.0]$. The contour map is given in Fig.~(\ref{fig:Fig0_1_b}), in grey color. 

\item On the other hand, when $z \in [0,2.5]$, the most likely best fit values with $1\sigma$ error are $H_{0} = 67.82 \pm 0.15$ km/s/Mpc, $ \alpha = 0.569\pm 0.002$ and $\gamma = 0.215 \pm 0.007$. The contour map is given in Fig.~(\ref{fig:Fig0_1_b}), in red color. 

\item Eventually, when $z \in [0,5.0]$, the most likely best fit values with $1\sigma$ error are $H_{0} = 68.03 \pm 0.15$ km/s/Mpc, $ \alpha = 0.574 \pm 0.002$ and $\gamma = 0.229 \pm 0.003$. The contour map is given in Fig.~(\ref{fig:Fig0_1_b}), in blue color. 

\end{itemize}

\begin{figure}[t!]
 \begin{center}$
 \begin{array}{cccc}
\includegraphics[width=80 mm]{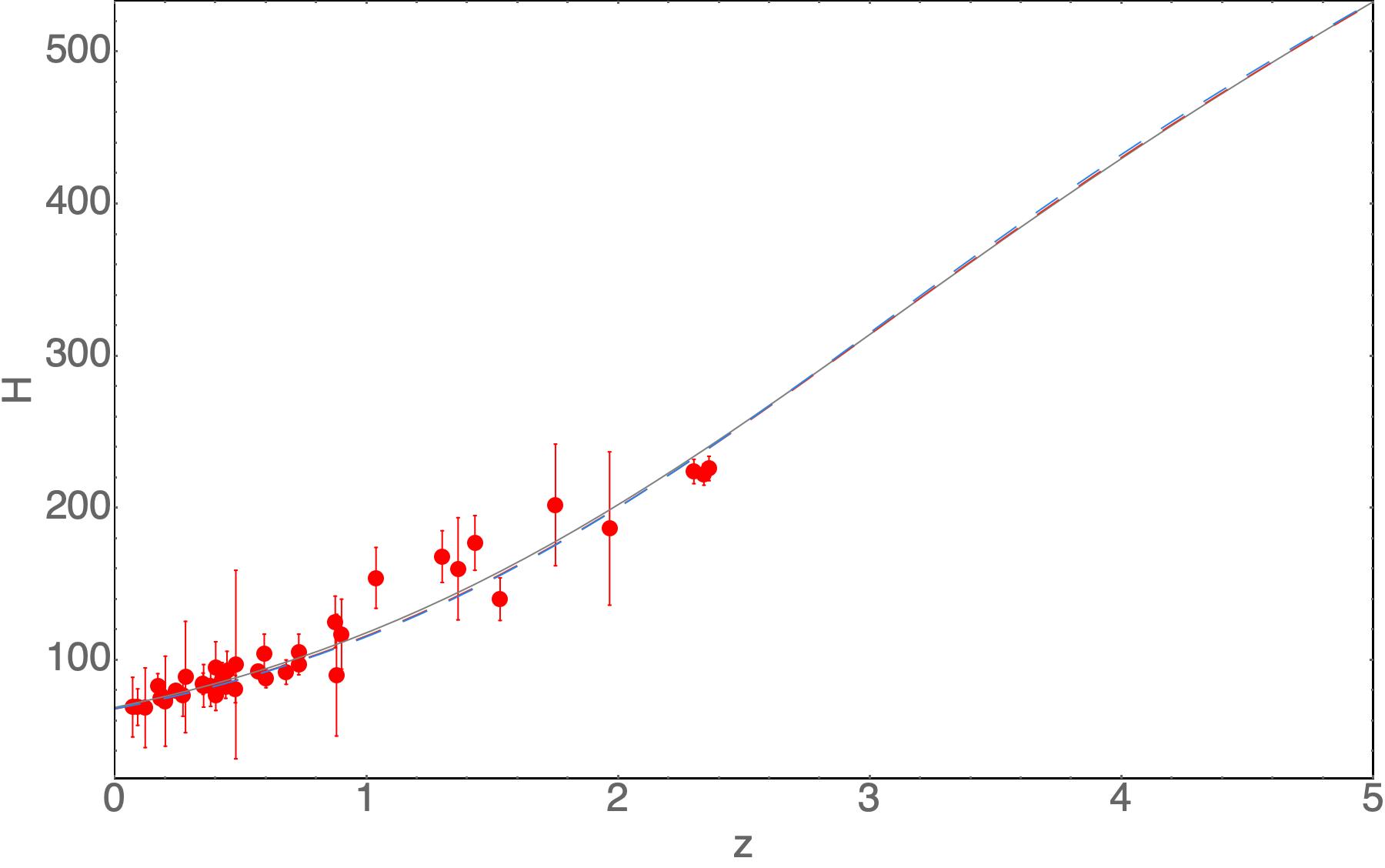}&&
\includegraphics[width=80 mm]{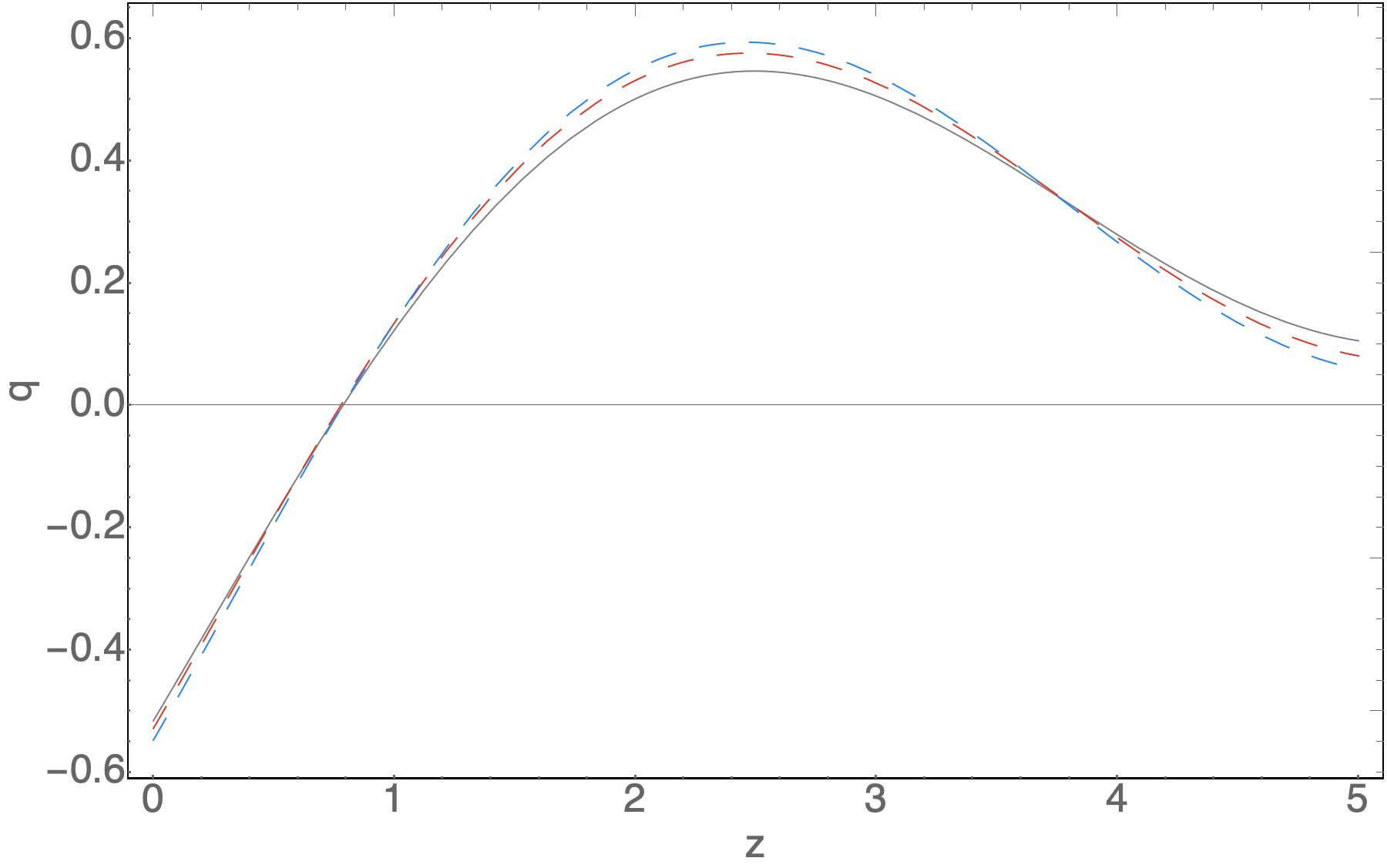}\\
 \end{array}$
 \end{center}
\caption{The behavior of the Hubble parameter in comparison with known $H(z)$ data is depicted on the left-hand side of the figure. The grey curve is a plot of the Hubble parameter for the best-fit values of the model parameters, when $z \in [0,2]$, the dashed red curve corresponds to the case when $z \in [0,2.5]$, while the dashed blue curve describes the case when $z \in [0,5]$. The red dots represent the known observational $H(z)$ data as in Table~\ref{tab:Table0}. The right-hand side depicts the graphical behavior of the deceleration parameter $q(z)$. The model is given by Eqs.~(\ref{eq:Gc2}) and (\ref{eq:HGc2}). In both cases, only the best-fit values for the model-free parameters obtained by the Bayesian Learning approach and presented in Table~\ref{tab:Table2} have been used.}
 \label{fig:Fig0_2_b}
\end{figure}

Our conclusion is that this model cannot solve the $H_{0}$ tension problem. However, contrary to the first model considered, this one can effectively explain the high-redshift expansion rate data. This can be seen from the plot on the left-hand side of Fig.~(\ref{fig:Fig0_2_b}). The graphic behavior of the deceleration parameter $q(z)$ presented on the right-hand side of Fig.~(\ref{fig:Fig0_2_b}) shows that the model can explain the accelerated expansion of the Universe, too. It should be mentioned that we learned very tight constraints on the model parameters. Moreover, the mean values of the model parameters can be affected significantly, depending on the redshift range used to generate the learning expansion rate data. The model is interesting to explain the BOSS results for the expansion rate data at $z=2.34$.

The results obtained in this last case motivated us to consider another model, with $\Gamma_{c} = 3 \alpha H_{0} + 3 \beta \cos(1+z) (1+z)^{\gamma}H (z)$ matter creation rate. We have learned the constraints to be, in this case, $H_{0} = 71.98^{+0.18}_{-0.16}$ km/s/Mpc, $\alpha = 0.5172 \pm 0.0024$,  $\beta = 0.056 \pm 0.048$ and $\gamma = 0.081 \pm 0.006$, when $z\in[0,2]$, and $H_{0} = 71.94^{+0.19}_{-0.18}$ km/s/Mpc, $\alpha = 0.5156\pm 0.0025$,  $\beta = 0.099 \pm 0.054$ and $\gamma = 0.019 \pm 0.005$, when $z\in[0,2.5]$. We do not exhibit this model separately, because we soon learned that it should be rejected, too. Indeed, it is in huge tension with high-redshift expansion rate data. The model we shall next discuss, however, has interesting features that could be used to solve the $H_{0}$ tension problem. It will indicate how the initial form of the matter creation rate must be modified in order to make a viable cosmological model with a matter creation mechanism.

\subsection{Model with $\Gamma_{c} = 3 \alpha H_{0} + 3 \beta H(z)^{\gamma}$}

The last model to be considered here has the following matter creation rate

\begin{table}
  \centering
    \begin{tabular}{ | c | c | c | c | c | c |  p{2cm} |}
    \hline
    
$\Gamma_{c} = 3 \alpha H_{0} + 3 \beta H(z)^{\gamma}$ & $H_{0}$ & $\alpha$ & $\beta$ & $\gamma$ \\
      \hline
 
  when $z\in[0,2]$ & $73.57 \pm 0.15$ km/s/Mpc & $0.561 \pm 0.013$ &  $0.272^{+0.051}_{-0.047}$ & $0.597 \pm 0.042$\\
          \hline
          
 when $z\in[0,2.5]$ & $73.56 \pm 0.15$ km/s/Mpc & $0.571\pm 0.011$ &  $0.237 \pm 0.051$ &  $0.543 \pm 0.042$ \\
          \hline
          
when $z\in[0,5]$  & $73.03 ^{+0.15}_{-0.16}$ km/s/Mpc& $0.576^{+0.005}_{-0.004}$ &  $0.208 \pm 0.052$ &  $0.364\pm 0.044$ \\

           \hline
 
     \end{tabular}
\caption{Best fit values and $1\sigma$ errors estimated for the model given by Eq.~(\ref{eq:Gc3}), for $z \in [0,2.0]$, $z \in [0,2.5]$ and $z \in [0,5]$, respectively. The results have been obtained from our Bayesian (Probabilistic) Machine Learning approach, where the generative-based process relies on Eq.~(\ref{eq:HGc2}), using $H_{0}\in[65.5,85.5]$, $\alpha \in [-1.5, 1.5]$, $\beta \in [-0.5, 0.5] $ and $\gamma \in [-0.5,1.5]$ as flat priors. The analysis is based on 10 chains and in each chain, 15,000 ``observational'' data sets have been simulated/generated.}
  \label{tab:Table3}
\end{table} 

\begin{equation}\label{eq:Gc3}
\Gamma_{c} = 3 \alpha H_{0} + 3 \beta H(z)^{\gamma}.
\end{equation}

Once more, starting from Eqs.~(\ref{eq:FEQ1}) and (\ref{eq:FEQ2}) from one side, and taking also into account Eqs.~(\ref{eq:Pc}) and (\ref{eq:Gc3}), after some algebra, we obtain the expansion rate dynamics of the Universe, as follows
\begin{equation}\label{eq:HGc3}
H^{\prime}(z) = -\frac{3 \left(\beta  H(z)^{\gamma }-H(z)+\alpha  H_{0}\right)}{2 (z+1)},
\end{equation}
where the prime stands for the derivative with respect to the redshift. Imposing $H_{0}\in[65.5,85.5]$, $\alpha \in [-1.5, 1.5]$, $\beta \in [-0.5, 0.5] $ and $\gamma \in [-0.5,1.5]$ as flat priors on the free parameters, we have used $10$ chains to generate $15,000$ ``observational'' data-sets in each one. The learned constraints can be found in Table~\ref{tab:Table3}. In particular, we learned that

\begin{figure}[h!]
 \begin{center}$
 \begin{array}{cccc}
\includegraphics[width=120 mm]{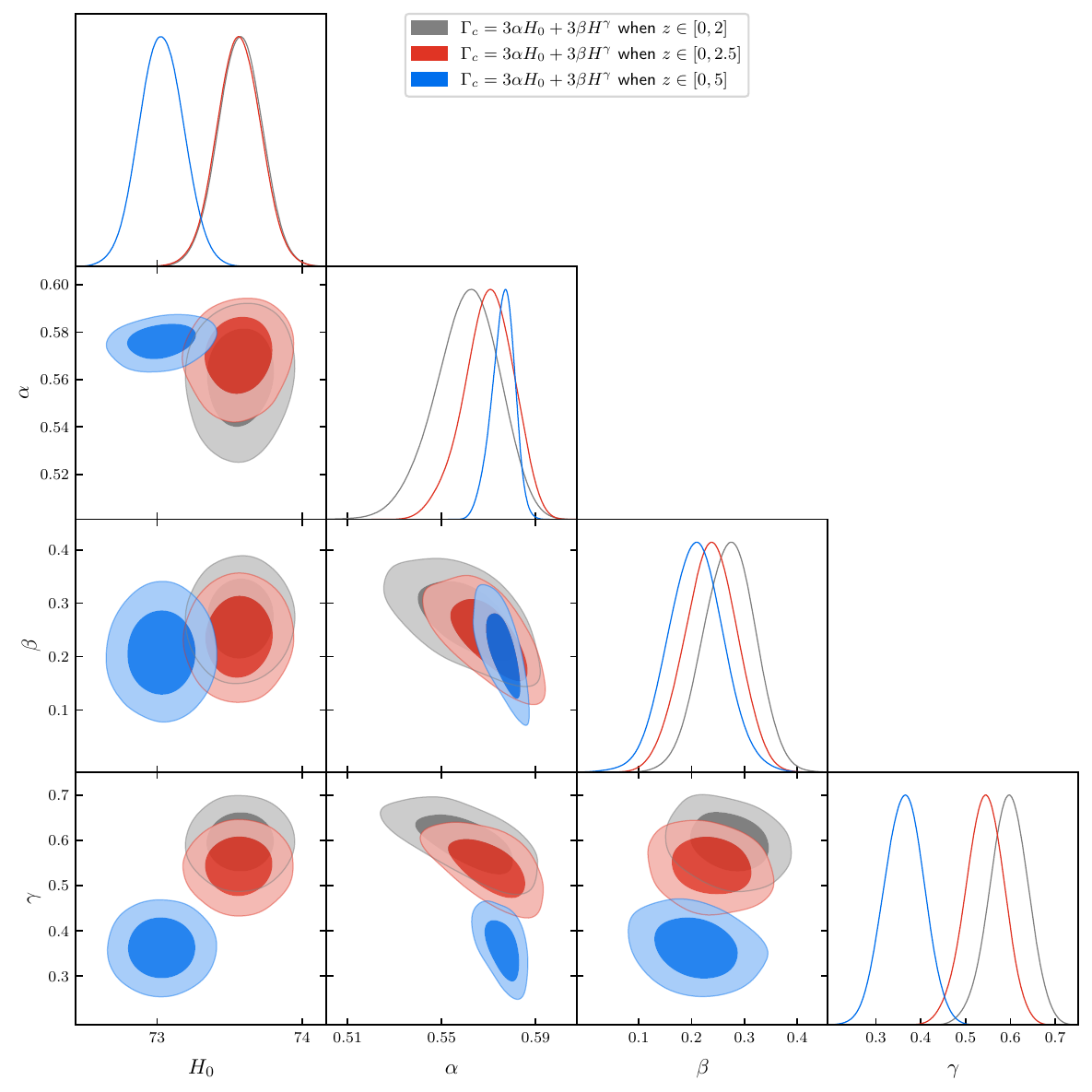}
 \end{array}$
 \end{center}
\caption{Contour maps of the model given by Eq.~(\ref{eq:Gc3}) for $z \in [0,2.0]$, $z \in [0,2.5]$ and $z \in [0,5]$, respectively. The best-fit values of the model parameters are presented in Table~\ref{tab:Table3}. In all three cases, $H_{0}\in[65.5,85.5]$, $\alpha \in [-1.5, 1.5]$, $\beta \in [-0.5, 0.5] $ and $\gamma \in [-0.5,1.5]$ as flat priors have been imposed to generate ``observational'' expansion rate data from Eq.~(\ref{eq:HGc3}). The analysis is based on 10 chains and in each chain, 15,000  ``observational'' data sets have been simulated/generated.}
 \label{fig:Fig0_1_c}
\end{figure}

\begin{itemize}

\item The best fit values for the model-free parameters with $1\sigma$ error are $H_{0} = 73.57 \pm 0.15$ km/s/Mpc, $\alpha = 0.561 \pm 0.013$, $\beta = 0.272^{+0.051}_{-0.047}$ and $\gamma = 0.597 \pm 0.042$. The contour map is given in Fig.~(\ref{fig:Fig0_1_c}), in grey color. 

\item On the other hand, when $z \in [0,2.5]$, the most likely best fit values with $1\sigma$ error are $H_{0} = 73.56 \pm 0.15$ km/s/Mpc, $\alpha = 0.571\pm 0.011$, $\beta = 0.237 \pm 0.051$ and $\gamma = 0.543 \pm 0.042$. The contour map is given in Fig.~(\ref{fig:Fig0_1_b}), in red color. 

\item To finish, when $z \in [0,5.0]$, the most likely best fit values with $1\sigma$ error are  $H_{0} = 73.03 ^{+0.15}_{-0.16}$ km/s/Mpc, $\alpha = 0.576^{+0.005}_{-0.004}$,  $\beta = 0.208 \pm 0.052$ and $\gamma = 0.364\pm 0.044$. The contour map is given in Fig.~(\ref{fig:Fig0_1_b}), in blue color. 

\end{itemize}

\begin{figure}[t!]
 \begin{center}$
 \begin{array}{cccc}
\includegraphics[width=80 mm]{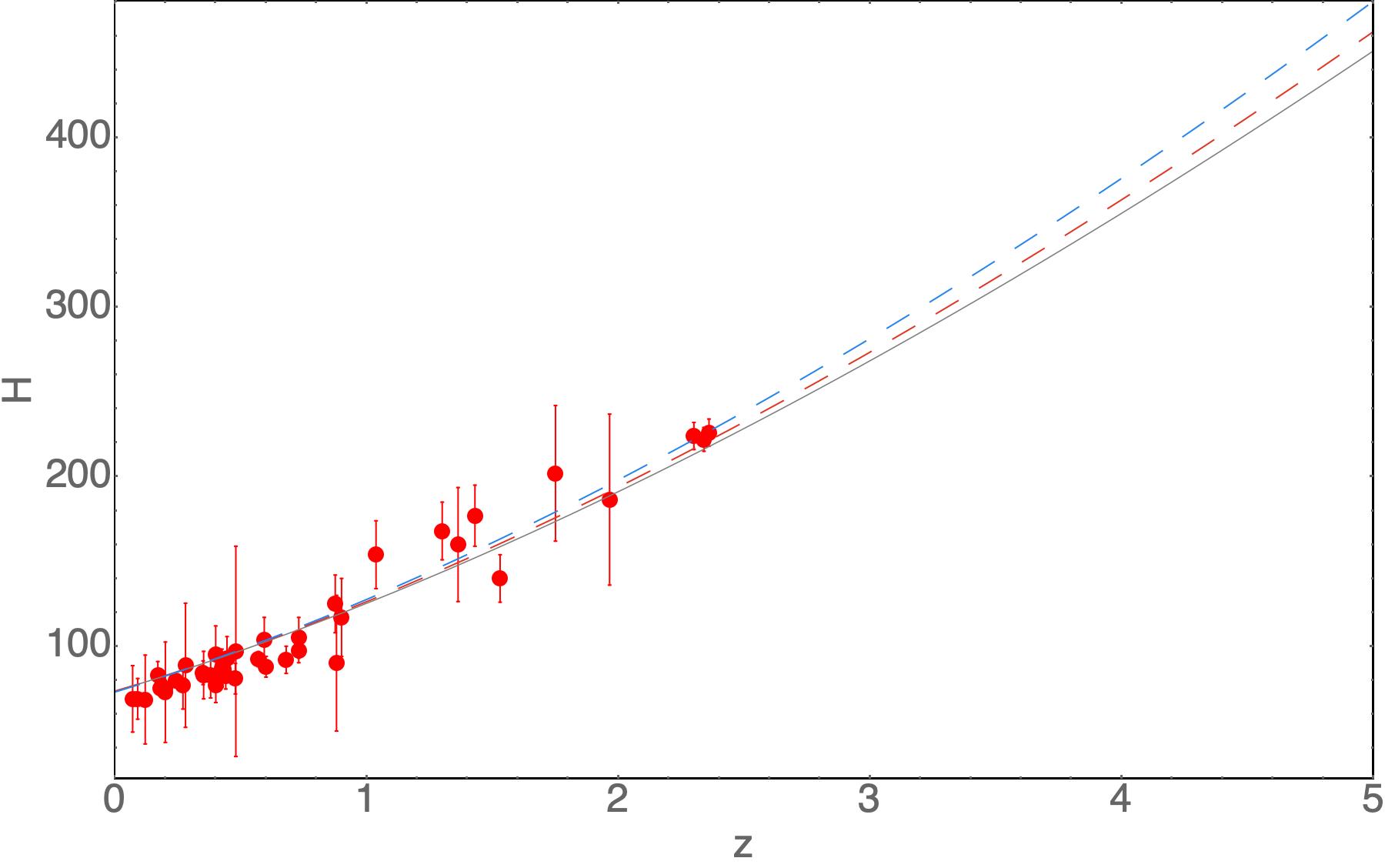}&&
\includegraphics[width=80 mm]{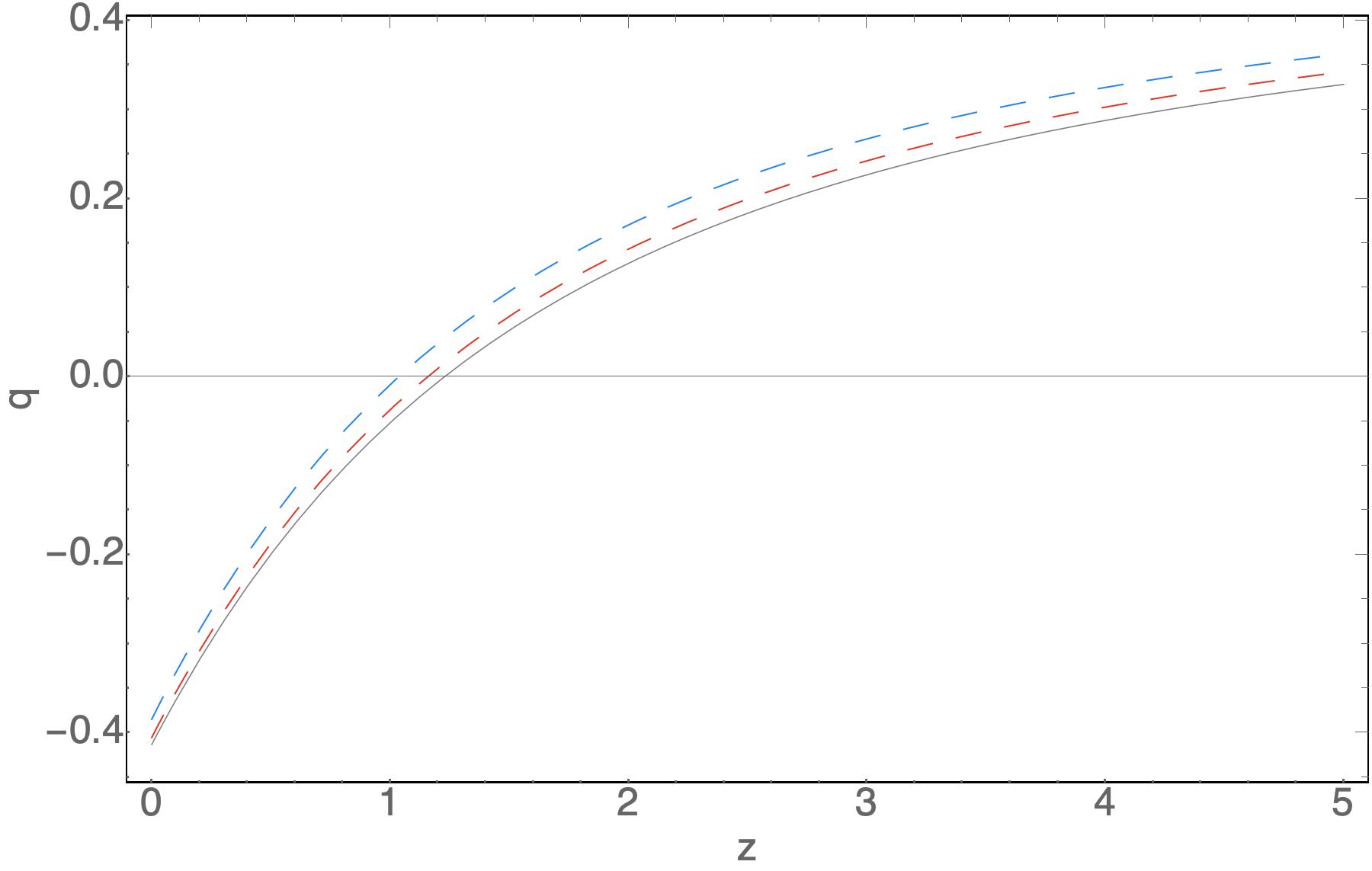}\\
 \end{array}$
 \end{center}
\caption{The graphic behavior of the Hubble parameter in comparison with known $H(z)$ data is depicted on the left-hand side. The grey curve is a plot of the Hubble parameter for the best-fit values of the model parameters, when $z \in [0,2]$, the dashed red curve corresponds to the case when $z \in [0,2.5]$, while the dashed blue curve is for $z \in [0,5]$. The red dots represent the known observational $H(z)$ data, as in Table~\ref{tab:Table0}. We plot the deceleration parameter $q(z)$ on the right-hand side. The model is given by Eqs.~(\ref{eq:Gc3}) and (\ref{eq:HGc3}). In both cases, only the best-fit values for the model-free parameters obtained by the Bayesian Learning approach and presented in Table~\ref{tab:Table3} have been used.}
 \label{fig:Fig0_2_c}
\end{figure}

The graphic behavior of the expansion rate given by Eq.~(\ref{eq:HGc3}) for learned-based values of the model-free parameters can be found on the left-hand side of Fig.~(\ref{fig:Fig0_2_c}). This plot shows that the model can indeed solve the $H_{0}$ tension problem and also that, reasonably well, it can explain the result of the BOSS. To end our discussion we turn to the right-hand side plot of Fig.~(\ref{fig:Fig0_2_c}), representing the graphical behavior of the deceleration parameter obtained for this model. It clearly points to an accelerated expansion of the late-time universe, thus confirming that particle creation can indeed be used to explain it.

\section{Conclusions}\label{sec:conc}

In this paper, a Bayesian (Probabilistic) Machine Learning approach has been used to ascertain whether the $H_{0}$ tension problem could be actually solved by employing a matter creation procedure. In the recent literature, matter creation mechanisms have been intensively studied since, for instance, they may lead to an explanation of the accelerated expansion of the universe without the need for dark energy. It is also crucial to understand whether or not they can be useful to solve, or at least alleviate, the $H_{0}$ tension problem, too. Removing dark energy, for one, can be very useful to find if the problem is related to some bias in the observational data. If we prove some success with the learning approach, then it could be further used to construct and constrain dark energy models and extensions of General Relativity. As a general point, it should be made clear that the way to choose a reasonable matter creation rate is here directly related to the observational data. 

Differently, as in previous studies, we have here used Machine Learning to learn if it is possible to construct reasonable matter creation rates interesting for cosmological applications. In the paper, we have used the expansion rate of the universe to be the observable generated during the learning process. Specifically, we have first constrained the models for the redshift ranges: $z\in [0,2]$~(cosmic chronometers) and $z\in [0,2.5]$~(cosmic chronometers + BAO), covering known $H(z)$ observations. In addition, we have constrained the models using data generated for the extended redshift range $z\in[0,5]$, taking into account that this range will not possibly be covered until results from future planned missions become available. Taking into account that there are numerous ways to construct the matter creation rate, we decided, at these initial stages of the learning, to use a specific family of models. In particular, we have studied models with $\Gamma_{c} = 3 \alpha H_{0} + F(H)$, where the matter creation rate can be parameterized in terms of $H_{0}$ and some function, $F(H)$, of the Universe expansion rate. 

The initial results obtained from the learning method look actually promising. They indicate that the Bayesian Machine Learning procedure is truly useful. In particular, when one considers a model with a matter creation rate of the form:
$\Gamma_{c} = 3 \alpha H_{0} + 3 \beta H (1+z)^{\gamma}$, with $\alpha$, $\beta$ and $\gamma$ the free parameters to be learned. We have found that this model can be useful to alleviate the $H_{0}$ tension problem. However, when the generated expansion rate data for $z\in[0,5]$ has been used to perform the learning, we proved that the associated model had to be rejected. This is due to a huge discrepancy between the available expansion rate data for $z\in [1.7, 2.5]$ and the theoretical results corresponding to the learned constraints. On the other hand, we have checked that the model with matter creation rate given by $\Gamma_{c} = 3 \alpha H_{0} + 3 \beta H$  cannot solve the $H_{0}$ tension problem, either. Even if this model can efficiently explain the low-redshift expansion rate data, there is still a huge discrepancy with the high-redshift expansion data. It should be also mentioned that the model with $\Gamma_{c} = 3 \alpha H_{0} + 3 \gamma H \cos(1+z)$, again, does not allow to solve the $H_{0}$ tension problem. Moreover, forecasting with $z\in[0,5]$ eventually gives a hint that the model has to be rejected, too. 

Remarkably, the model with $\Gamma_{c} = 3 \alpha H_{0} + 3 \beta H^{\gamma}$ is a viable one, with a view at solving the $H_{0}$ tension problem. Compared to the model given by Eq. (\ref{eq:GcOdintsov}), we see that the model provides a solution to the $H_{0}$ tension problem, mainly because of the $3 \alpha H_{0}$ term explicitly introduced into the matter-creation rate. Moreover, from the learned results, there is a strong hint that this specific model is very robust and will stand in good agreement with future high-redshift expansion rate data. 

The results of the Machine Learning approach used here indicate that, in all models, the study of the deceleration parameter shows that matter creation can be taken as a reasonable way to explain the accelerated expansion of the universe. The only obstacle is learning how to parametrize it correctly; with the help of our particular model, we could show which specific forms were to be rejected. Further analysis will also be necessary to ascertain how higher-order derivatives of $H(z)$ and the $3 \alpha H_{0}$ term may affect the results obtained with the learning method. 

Further research will be required to gain a better understanding of the impact of matter creation on cosmology. However, it was already very interesting to see how Machine Learning can be applied with the ultimate goal of obtaining the most general form of the matter creation rate, which may be universally valid for cosmological purposes. In this regard, the idea is to find a Gaussian Process that can be used for these purposes, a question that we expect to address in future work. It should be emphasized, though, that the proposed matter creation model will heavily rely on the quality of the data. Additionally, existing correlations and biases between different observational datasets can significantly impact the reconstruction process. This is an important issue to study, as recent literature in AI and ML has shown that considering these factors in Gaussian Processes can greatly influence the results obtained. This may not have a significant impact on cosmological studies currently, as observational data is not yet abundant enough to necessitate such methods. However, it remains an important consideration for the future. This paper can be viewed as an initial step in attempting to integrate Machine Learning with matter creation cosmology, and several intriguing questions have already been raised, despite the limitations being faced. We hope to achieve, through this effort, a more comprehensive and unbiased understanding of how and why the matter creation mechanism operates in the cases under consideration.

Furthermore, our previous research indicates that Bayesian Machine Learning is a powerful tool to utilize cosmic opacity, swampland, and strong gravitational lensing physics to constrain cosmological models and provide a solution to the H0 tension problem. We are confident that the angular scale of the sound horizon will be another valuable tool (see \cite{tur_2} and references therein) used with Bayesian Machine Learning to confirm the hints we have discovered in our recent paper. This topic remains to be studied and reported in forthcoming papers along with the other questions mentioned above.

\section*{Acknowledgement}
We appreciate the valuable discussion and comments from the referee, which have greatly improved the manuscript.
This work has been partially supported by MICINN (Spain), project PID2019-104397GB-I00, of the Spanish State Research Agency program AEI/10.13039/501100011033, by the Catalan Government, AGAUR project 2021-SGR-00171, and by the program Unidad de Excelencia María de Maeztu CEX2020-001058-M. M.K. has been supported by a Juan de la Cierva-incorporación grant (IJC2020-042690-I).


\begin{thebibliography}{1}

\bibitem{Riess}
A.G. Riess et al., [Supernova Search Team], Astron. J. 116, 1009 (1998).

\bibitem{Ade1}
P.A.R. Ade et al. [Planck Collaboration], Astron. Astrophys. 571, A16 (2014).

\bibitem{Ade2}
P.A.R. Ade et al. [Planck Collaboration], A $\&$ A 594, A13 (2016).

\bibitem{Ade3}
N. Aghanim et al. [Planck Collaboration], arXiv:1807.06209.

\bibitem{Riess2}
A.G. Riess et al., Ap J 861, 126 (2018).

\bibitem{Perlmutter}
S. Perlmutter et al.,  [Supernova Cosmology Project Collaboration], Astrophys. J. 517, 565 (1999).

\bibitem{Alam}
S. Alam et al. [BOSS Collaboration], Mon. Not. Roy. Astron. Soc. 470, no.3, 2617 (2017).

\bibitem{Troxel}
M.A. Troxel et al. [DES Collaboration], Phys. Rev. D 98, no.4, 043528 (2018).

\bibitem{Hinshaw}
G. Hinshaw et al. [WMAP Collaboration], Astrophys. J. Suppl. 208, 19 (2013).

\bibitem{Plank}
N. Aghanim et al. (Planck Collaboration),  arxiv:1807.06209.

\bibitem{Hubble}
A. G. Riess et al., Ap  J 861, 126 (2018).

\bibitem{Hubble1}
A. G. Riess et al., ApJL 934 L7 (2022).

\bibitem{Perivolaropoulos_LCDM}
L. Perivolaropoulos, F. Skara, New Astronomy Reviews, Volume 95, 2022, 101659.

\bibitem{Referee_1}
E. Abdalla et al,  J. High En. Astrophys. 2204, 002 (2022).

\bibitem{IDE_1}
S. Kumar, R. C. Nunes, Phys. Rev. D 96, 103511 (2017).

\bibitem{IDE_2}
E. Di Valentino, A. Melchiorri, O. Mena, Phys. Rev. D 96, 043503 (2017).

\bibitem{IDE_3}
S. Pan et al., Phys. Rev. D 100, 103520 (2019). 

\bibitem{IDE_4}
E. Di Valentino et al, Phys. Rev. D 101, 063502 (2020).

\bibitem{IDE_5}
E. Di Valentino et al, Phys. Dark Univ. 30, 100666 (2020).

\bibitem{IDE_6}
A. Bernui et al, Phys. Rev. D 107, 103531 (2023).

\bibitem{IDE_7}
L. A. Escamilla et al, JCAP 11, 051 (2023).

\bibitem{Elizalde_9}
E. Elizalde et al, Phys.Rev.D 102 (2020) 12, 123501.

\bibitem{Elizalde_8}
E. Elizalde, M. Khurshudyan, Phys. Dark Univ. 37 (2022) 101114.

\bibitem{Elizalde_7}
E. Elizalde, M. Khurshudyan, Eur. Phys. J. C 81 (2021) 4, 335, Eur. Phys. J. C 81 (2021) 5, 438 (erratum).

\bibitem{Elizalde_H0}
E. Elizalde, J. Gluza, M. Khurshudyan, arXiv:2104.01077.

\bibitem{Elizalde_1}
E. Elizalde, M. Khurshudyan, Phys. Rev. D 99 (2019) 10, 103533.

\bibitem{Elizalde_2}
E. Elizalde, M. Khurshudyan, Eur. Phys. J. C 82 (2022) 9, 81.

\bibitem{Elizalde_3}
E. Elizalde et al, arxiv:2203.06767.

\bibitem{Elizalde_4}
M. Aljaf et al, Eur. Phys. J. C 82 (2022) 12, 1130.

\bibitem{Elizalde_5}
K. Yerzhanov et al,  Mod. Phys. Lett. A 36 (2021) 31, 2150222.

\bibitem{Elizalde_6}
M. Aljaf et al, Int. J. Mod. Phys. A 37 (2022) 34, 2250211.

\bibitem{Elizalde_H0_recent}
M. Khurshudyan, E. Elizalde, Constraints on prospective deviations from the cold dark matter model using a Gaussian Process, submitted (2023).

\bibitem{Elizalde_10}
Y.F. Cai et al, Astrophys.J. 888, 62 (2020).

\bibitem{MK1}
E. Elizalde et al, Int. J. Mod. Phys. D 28, No. 01, 1950019 (2019).

\bibitem{MK13}
M. Khurshudyan, R. Myrzakulov, Eur. Phys. J. C 77: 65 (2017).  

\bibitem{MK12}
C. Li et al, Phys. Lett. B 80, 135141(2020).

\bibitem{Nojiri:2022ski}
S.~Nojiri et al,  Nucl. Phys. B 980, (2022), 115850.

\bibitem{Odintsov:2022eqm}
S.~D.~Odintsov and V.~K.~Oikonomou, EPL 137, (2022) no.3, 39001.

\bibitem{Nojiri:2021dze}
S.~Nojiri et al, Phys. Dark Univ. 32, (2021), 100837.

\bibitem{Odintsov:2020qzd}
S.~D.~Odintsov et al, Nucl. Phys. B 966 (2021), 115377.

\bibitem{Odintsov:2017icc}
 S.D. Odintsov et al., Phys. Rev. D 96, no.4, 044022 (2017).
 
 \bibitem{Bamba:2012cp}
K. Bamba et al, Astrophys. Space Sci. 342, 155 (2012). 
 
\bibitem{Mk17}
M. Aljaf et al, Eur. Phys. J. C  80:112 (2020).

\bibitem{Mk18}
E. Sadri et al, Eur. Phys. J. C  80:393 (2020).

\bibitem{Krishnan}
C. Krishnan et al, Phys. Rev. D 103, 103509 (2021).

\bibitem{Valentino}
E. Di Valentino et al, Class. Quantum Grav. 38 153001 (2021).

\bibitem{Jian-Ping}
Jian-Ping Hu and Fa-Yin Wang, Universe 2023, 9, 94.

\bibitem{Colgain}
E.O Colgain et al, Phys. Rev. D 106, L041301 (2022).

\bibitem{Colgain1}
E.O Colgain et al, arXiv:2206.11447.

\bibitem{Colgain2}
M. Malekjani et al, arXiv:2301.12725. 

\bibitem{H0End}
R. C. Nunes, JCAP 05, 052 (2018).

\bibitem{INStart}
I. Brevik et al, Int. J. Geom. Meth. Mod. Phys. 14, 1750185 (2017).

\bibitem{INStart_4}
S. Nojiri, S.D. Odintsov, Phys. Rev. D 72, 023003 (2005).

\bibitem{INStart_7}
I. Brevik et al, Int. J. Mod. Phys. D 26, no.14, 1730024 (2017).

\bibitem{INStart_9}
S.D. Odintsov et al, Annals Phys. 398, 238-253  (2018).

\bibitem{INEnd}
S.D. Odintsov et al,  Phys. Rev. D 101, 044010 (2020).

\bibitem{GP_0}
X. Rin et al, Astrophys. J. 932 (2022) 2, 131.

\bibitem{GP_1}
Peng-Ju Wu et al, arxiv:2209.08502.

\bibitem{GP_2}
J.L. Said et al, JCAP 06, 015  (2021).

\bibitem{GP_4}
S. Dhawan et al., Mon. Not. Roy. Astron. Soc. 506, L1 (2021).

\bibitem{GP_7}
R.C. Bernardo, J.L. Said,  JCAP 08, 027 (2021).

\bibitem{MC_start}
M.O. Calvao et al, Phys. Lett. A 162, 223 (1992).

\bibitem{MC_1}
J.A.S. Lima, A.S. Germano, Phys. Lett. A 170, 373 (1992).

\bibitem{MC_2}
J.A.S. Lima, J.S. Alcaniz, Astron. Astrophys. 348, 1 (1999).

\bibitem{MC_3}
C.P. Singh, A. Beesham, Astrophys. Space Sci. 336, 469 (2011).

\bibitem{MC_4}
Q. Yaun, J. Tong, Y. Ze-Long, Astrophys. Space Sci. 311, 407 (2007).

\bibitem{MC_5}
L.R.W. Abramo, J.A.S. Lima, Class. Quantum Gravity 13, 2953 (1996).

\bibitem{MC_6}
J.A.S. Lima et al, JCAP 10, 042 (2014).

\bibitem{MC_7}
R.O. Ramos et al, Phys. Rev. D 89, 083524 (2014).

\bibitem{MC_8}
J. de Haro, S. Pan, Class. Quantum Gravity 33, 165007 (2016).

\bibitem{tur}
O. Akarsu, N. M. Uzun, Phys. Dark Univ. 40, 101194 (2023).

\bibitem{MC_9}
S. Pan, S. Chakraborty, Adv. High Energy Phys. 201, 654025 (2015).

\bibitem{MC_10}
V.H. Cárdenas et al, Phys. Rev. D 101, 083530 (2020).

\bibitem{MC_end}
S. Pan et al, Mon. Not. R. Astron. Soc. 460, 1445 (2016).

\bibitem{MC_Ref_1}
H. Bondi and T. Gold, MNRAS 108, 252 (1948)

\bibitem{MC_Ref_2}
 F. Hoyle, MNRAS 108, 372 (1948)
 
\bibitem{HTable_0}
C. Zhang et al, Research in Astronomy and Astrophysics 14, 1221 (2014). 

\bibitem{HTable_0_Ref}
M. Moresco et al, Living Rev Relativ 25, 6 (2022). 

\bibitem{HTable_1}
M. Moresco et al, JCAP 05, 014 (2016).

\bibitem{HTable_2}
R. Jimenez et al,  Astrophys. J. 593, 622 (2003).

\bibitem{HTable_3}
D. Stern et al, JCAP 008 (2010).

\bibitem{HTable_4}
M. Moresco et al, JCAP 08, 006 (2012).

\bibitem{HTable_5}
J. Simon et al, Phys. Rev. D 71, 123001 (2005).

\bibitem{HTable_6}
M. Moresco, Mon. Not. R. Astron. Soc. Lett. 450, L16 (2015).

\bibitem{HTable_7}
E. Gaztanaga et al, Mon. Not. R. Astron. Soc. 399, 1663 (2009).

\bibitem{HTable_8}
C. Blake et al, Mon. Not. R. Astron. Soc. 425, 405 (2012).

\bibitem{HTable_9}
X. Xu et al, Mon. Not. R. Astron. Soc. 431, 2834 (2013).

\bibitem{HTable_10}
T. Delubac et al, Astronomy $\&$ Astrophysics 552, A96 (2013).

\bibitem{HTable_11}
T. Delubac et al, Astronomy $\&$ Astrophysics 574, A59 (2015).

\bibitem{HTable_12}
L. Samushia et al, Mon. Not. R. Astron. Soc. 429, 1514 (2013).

\bibitem{HTable_13}
A. Font-Ribera et al, JCAP 05, 027 (2014).

\bibitem{KL}
S. Kullback, R.A. Leibler,  Ann. Math. Stat. 22, 79–86 (1951).

\bibitem{ELBO}
D.M. Blei et al,  J. Am. Stat. Assoc. 112(518), 859–877 (2017). 

\bibitem{51}
A. Graves, Advances in Neural Information Processing Systems, pp. 2348-2356 (2011).

\bibitem{52}
N. Metropolis et al, J. Chem. Phys. 21, 1087–1092 (1953). 

\bibitem{53}
J. Regier et al, arXiv:1803.00113

\bibitem{54}
G. Gunapati et al, Publ. Astron. Soc. Austral. 39, e001 (2022). 

\bibitem{PyMC3}
J. Salvatier, T. Wiecki, C. Fonnesbeck, PeerJ Comput. Sci. 2, e55 (2016).

\bibitem{tur_2}
O. Akarsu et al, Phys. Rev. D 107, 123526 (2023).

\end{thebibliography}
\end{document}